\documentclass[letterpaper,12pt]{article}
\usepackage{color,amssymb,enumerate,float}
\usepackage{amsmath}

\usepackage{ifpdf}
\ifpdf
	\usepackage[pdftex]{graphicx}
	\usepackage[pdftex,unicode,implicit]{hyperref}

	\hypersetup{
  	pdftitle     = {}, 
  	pdfkeywords  = {},
  	pdfauthor    = {},
  	pdfcreator   = {pdf\LaTeXe\ with package \flqq hyperref\frqq},
  	pdfproducer  = {pdf\LaTeXe\ with package \flqq hyperref\frqq},
  	pdfpagemode  = UseNone,  
  	pdffitwindow = true,  
  	unicode      = true,
  	plainpages   = true,
  	colorlinks   = true,  
  	citecolor    = black,  
  	urlcolor     = blue, 
  	linkcolor    = black
	}

\else

  \usepackage[dvips]{graphicx}

\fi 

\makeatletter
\@addtoreset{equation}{section}
\makeatother

\begin{document}
	
\thispagestyle{empty}

\begin{center}
{\bf \LARGE Effective action of the Ho\v{r}ava theory: Cancellation of divergences}
\vspace*{15mm}

{\large Jorge Bellor\'{\i}n}$^{1,a}$,
{\large Claudio B\'orquez$^{2,b}$}
{\large and Byron Droguett}$^{1,c}$
\vspace{3ex}

$^1${\it Department of Physics, Universidad de Antofagasta, 1240000 Antofagasta, Chile.}
´
$^2${\it Facultad de Ingenier\'{i}a, Arquitectura y Diseño, Universidad San Sebasti\'{a}n, Lago Panguipulli 1390, Puerto Montt, Chile.}

$^a${\tt jorge.bellorin@uantof.cl,} \hspace{1em}
$^b${\tt claudio.borquez@uss.cl,} \hspace{1em}
$^c${\tt byron.droguett@uantof.cl}

\vspace*{15mm}
{\bf Abstract}
\begin{quotation}{\small
We compute the one-loop effective action of the Ho\v{r}ava theory, in its nonprojectable formulation. We take the quantization of the $(2+1)$-dimensional theory in the Batalin-Fradkin-Vilkovisky formalism, and comment on the extension to the $(3+1)$ case. The second-class constraints and the appropriate gauge-fixing condition are included in the quantization. The ghost fields associated with the second-class constraints can be used to get the integrated form of the effective action, which has the form of a Berezinian. We show that all irregular loops cancel between them in the effective action. The key for the cancellation is the role of the ghosts associated with the second-class constraints. These ghosts form irregular loops that enter in the denominator of the Berezinian, eliminating the irregular loops of the bosonic nonghost sector. Irregular loops produce dangerous divergences; hence their cancellation is an essential step for the consistency of the theory. The cancellation of this kind of divergences is in agreement with the previous analysis done on the $(2+1)$ quantum canonical Lagrangian and its Feynman diagrams.		
} 
	
\end{quotation}
\end{center}

\thispagestyle{empty}

\newpage

\section{Introduction}
The effective action is a very important object in a quantum theory. It is very useful to study the renormalizability of field theories. Indeed, the Slavnov identities can be established in terms of it, such that one may obtain relations for the poles and finite parts. Since correlation or Green functions can be conveniently derived from the effective action, the renormalization group flow can be applied to it. Moreover, for potentials with spontaneous symmetry breaking, the quantum dynamics naturally leads to the usage of effective fields and their effective action. The program of renormalization based on the effective action can be applied to the Ho\v{r}ava theory of quantum gravity \cite{Horava:2009uw}, whose proof of renormalizability in the case of the nonprojectable version is still pending.

The quantization of the nonprojectable case of the Ho\v{r}ava theory is a delicate issue since it is a theory with second-class constraints. On the other hand, the case of the projectable theory shows that a particular gauge-fixing condition is required \cite{Barvinsky:2015kil}. This gauge-fixing condition is noncanonical in the context of the basic phase space. In this case, the Batalin-Fradkin-Vilkovisky (BFV) formalism of quantization \cite{Fradkin:1975cq,Batalin:1977pb} is appropriate. For the nonprojectable theory, the measure of the second-class constraints \cite{Senjanovic:1976br,Fradkin:1977xi} must be incorporated. We have studied this problem in previous analysis \cite{Bellorin:2019gsc,Bellorin:2021tkk,Bellorin:2021udn}, including the projectable case, obtaining that the Lagrangian quantization of the projectable case with the appropriate nonlocal gauge-fixing condition can be obtained from the BFV quantization. With the BFV quantization we have also studied the Becchi-Rouet-Stora-Tyutin (BRST) symmetry of the nonprojectable case rigorously \cite{Bellorin:2022efu}, which is an essential aspect for renormalization.

In this paper we obtain the one-loop effective action of the Ho\v{r}ava theory based on the BFV quantization. We focus on the nonprojectable case, which is a more general version that we expect it is closer to general relativity in the limit of large distances. To present the BFV quantization, we take the case of the $(2+1)$-dimensional theory, which was carried out in Refs.~\cite{Bellorin:2021udn,Bellorin:2022qeu} with the required gauge-fixing condition. We discuss the extension to the $(3+1)$ case, which is analogous to the $(2+1)$ case, due to the same functional form of the effective action and the same classification between regular and irregular propagators. In the common case of a field theory, the one-loop effective action with bosonic fields is given in terms of the determinant of the matrix of second derivatives of the action. In the case of the Ho\v{r}ava theory, one has fermionic fields that play the role of ghost fields needed for the quantization. There are two kinds of these fields: the BFV ghosts, associated with the gauge symmetry, and the ghosts associated with the measure of the second-class constraints. We find that all these ghosts can be used to arrive at an integrated form of the effective action, which is given in terms of the Berezinian of the matrix of second derivatives of the action. The Berezinian is a generalization of the determinant. It is well known in supersymmetric theories \cite{Witten:2012bg,Neufeld:1998js}.

Our second goal in this study is to use the integrated form of the effective action to analyze the cancellation of certain kind of divergences arising in the loop expansion. We call them divergences coming from irregular loops. This characterization is related to the problem of the gauge fixing in the original proof of renormalization of the projectable case \cite{Barvinsky:2015kil,Barvinsky:2017zlx}. In the projectable theory, the gauge-fixing condition chosen leads to regular propagators for all quantum fields. We give the definition of regular propagator in the body of the paper. The regular form is crucial to get the renormalization; this is the origin of the nonlocal gauge-fixing condition used on the side of the Lagrangian quantization. It turns out that, in the case of the nonprojectable theory, the analogous gauge-fixing condition can be imposed on the side of the BFV quantization, but the auxiliary fields associated with the measure of the second-class constraints acquire irregular propagators. Moreover, irregular loops formed completely with these irregular propagators produce a globally multiplicative divergence; hence this divergence cannot be subtracted by counterterms in a meaningful way. This is the reason why the irregular loops of the Ho\v{r}ava theory are dangerous.

In the previous analysis done in Ref.~\cite{Bellorin:2022qeu}, we studied the issue of the irregular propagators present in the theory. We performed the analysis directly on the Feynman diagrams derived from the quantum BFV formalism. We found that all the irregular loops cancel exactly between them. Specifically, there are two kinds of fields associated with the second-class constraints: a Lagrange multiplier and some ghost fields. They have the same irregular propagators, and the cancellation holds thanks to the exact matching between bosonic--fermionic loops. Various topological results from the diagrams of the theory and the structure of the quantum Lagrangian support this matching \cite{Bellorin:2022qeu}. The cancellation of the dangerous irregular loops is essential for the survival of the theory. All the remaining loops, at all orders, are regular.

In this paper we study the irregular loops, now in the framework of the effective action that we obtain. It is notorious the functional dependence of the effective action on the ghosts associated with the second-class constraints: they arise in the denominator of the Berezinian. Due to this, we are motivated to look for the cancellations within the effective action. This is a very important step for the renormalization program. The expectation is that the effective action can be used to control the poles; hence a necessary previous step is the elimination of the dangerous irregular loops.

The effective action of the Ho\v{r}ava theory has been studied previously by other authors, specifically for the projectable version. In Ref.~\cite{Benedetti:2013pya}, analysis of the renormalization group flow were done on a toy model in $(2+1)$ dimensions, obtaining preliminary results about the asymptotic freedom of the theory. Those authors performed the heat kernel expansion. The complete analysis of the $(2+1)$ projectable case was done in Ref.~\cite{Barvinsky:2017kob}, showing the asymptotic freedom property of the theory. The renormalization group flow for the projectable theory in $(3+1)$ dimensions has been studied in Refs.~\cite{Barvinsky:2019rwn,Barvinsky:2021ubv,Barvinsky:2023uir}. In particular, in Ref.~\cite{Barvinsky:2023uir} asymptotically-free fixed points were found that are connected by trajectories to the region where the low-energy coupling constant gets the relativistic form. Non-monotonic behavior arises in the flow. Previous studies on the effective action and the renormalization group flow in the projectable version can be found in \cite{Contillo:2013fua,D'Odorico:2014iha,D'Odorico:2015yaa,Griffin:2017wvh}.

This paper is organized as follows. In section 2 we collect previous results on the BFV quantization of the theory, the appropriate gauge fixing and the propagators of the quantum fields. The exposition is based on the $(2+1)$ theory, showing at the end the required ingredients for the quantization of the $(3+1)$ theory. In section 3 we obtain the one-loop effective action, and we show the cancellation of irregular loops. Finally we present some conclusions.


\section{BFV Quantization}
In this section we summarize previous results about the quantization of the Ho\v{r}ava theory \cite{Bellorin:2021tkk,Bellorin:2021udn, Bellorin:2022qeu}.

\subsection{Path integral}
The fields representing the gravitational interaction are the ADM variables $N(t,\vec{x})$, $N^i(t,\vec{x})$ and $g_{ij}(t,\vec{x})$. We develop the analysis for the case of $d=2$ spatial dimensions. At the end we comment on the $d=3$ case. This theory exhibits two cases: projectable and nonprojectable case. The last one is defined by the condition of the lapse function $N$ can be a function of the time and the space, whereas in the projectable case it is restricted to be a function only of time. We study the nonprojectable case. Given the foliation, the classical action of the nonprojectable theory is \cite{Horava:2009uw,Blas:2009qj}
\begin{equation}
 S = \int dt d^2x \sqrt{g} N \left( K_{ij} K^{ij} - \lambda K^2 - \mathcal{V} \right) \,,
\end{equation}
where the kinetic terms are defined in terms of the extrinsic curvature tensor,
\begin{equation}
 K_{ij} = \frac{1}{2N} \left( \dot{g}_{ij} - 2 \nabla_{(i} N_{j)} \right) \,,
 \qquad
 K = g^{ij} K_{ij} \,.
\end{equation}
$\mathcal{V}$ is called the potential. It contains the higher-order spatial derivatives, whose order is determined with the anisotropic parameter $z$, such that the higher-order terms are of $2z$ order in spatial derivatives. In the $(2+1)$-dimensional case, which requires a potential of $z=2$ order for power-counting renormalizability, the complete potential is \cite{Sotiriou:2011dr}
\begin{eqnarray}
\mathcal{V} &=&
-\beta R-\alpha a^{2}+\alpha_{1}R^{2}+\alpha_{2}a^{4}+\alpha_{3}R a^{2}+\alpha_{4}a^{2}\nabla_{i}a^{i}
\nonumber \\ &&
+\alpha_{5} R\nabla_{i}a^{i} 
+\alpha_{6}\nabla^{i}a^{j}\nabla_{i}a_{j}+ \alpha_{7}(\nabla_{i}a^{i})^{2} \,,
\label{potencial}
\end{eqnarray}
where 
\begin{equation}
a_i = \frac{\partial_i N}{N} \,.
\end{equation} 
$\nabla_i$ and $R$ are the covariant derivative and the Ricci scalar of the spatial metric $g_{ij}$, and the coupling constants of the theory are $\lambda$, $\beta$, $\alpha$, $\alpha_1$,...,$\alpha_7$.

In the canonical formalism \cite{Kluson:2010nf,Donnelly:2011df,Bellorin:2011ff}, the primary classical Hamiltonian results
\begin{equation}
	H_0 =
	\int d^{2}x
	\sqrt{g} N \left( \frac{\pi^{ij}\pi_{ij}}{g} 
	+ \bar{\sigma} \frac{\pi^{2}}{g} + \mathcal{V} \right) \,,
	\label{H0}
\end{equation}
where the canonically conjugate pairs are $(g_{ij},\pi^{ij})$ and $N$ with its conjugate momentum. This canonical momentum is zero due to the constraints of the theory; we discard it from the phase space.  We use the combinations of constants,
\begin{equation}
\sigma = \frac{1-\lambda}{1-2\lambda} \,,
\quad
\bar{\sigma} = \frac{\lambda}{1-2\lambda} \,,
\quad
\alpha_{67} = \alpha_6 + \alpha_7 \,.
\end{equation}  

The constraint that corresponds to the involutive functions under Dirac brackets in the BFV quantization is  
\begin{equation}
	\mathcal{H}_i =
	- 2 g_{ij} \nabla_k \pi^{kj} = 0 \,.
	\label{momentumconts}
\end{equation}
This constraint is the generator of the spatial diffeomorphism on the canonical pair $( g_{ij} , \pi^{ij} )$. The second-class constraints of the theory are the vanishing of the momentum conjugate to $N$, which we have already considered as a solved constraint, and
\begin{eqnarray}
	\theta_{1} &=&  
	\sqrt{g} N \left( \frac{\pi^{ij}\pi_{ij}}{g} 
	+ \bar{\sigma} \frac{\pi^{2}}{g} + \mathcal{V} \right) 
	+ \sqrt{g} \Big( 
	  2\alpha\nabla_{i}(Na^{i})
	- 4\alpha_{2}\nabla_{i}(Na^{2}a^{i})
	- 2\alpha_{3}\nabla_{i}(NRa^{i})
	\nonumber\\ &&
	+ \alpha_{4} \left( \nabla^{2}(Na^{2}) 
	-2\nabla_{i}( N a^{i} \nabla_{j} a^{j} ) \right)
	+ \alpha_{5}\nabla^{2}(NR)
	+ 2\alpha_{6}\nabla^{i}\nabla^{j}(N\nabla_{j}a_{i})
	\nonumber \\ &&
	+ 2\alpha_{7}\nabla^{2}(N\nabla_{i}a^{i})
	\Big) 
    = 0 \,.
	\label{theta2}
\end{eqnarray}
The primary Hamiltonian (\ref{H0}) is equivalent to the integral of this second-class constraint,
\begin{equation}
	H_{0} = \int d^{2}x \,\theta_{1} \,.
	\label{primaryhamiltonian}
\end{equation}

The BFV quantization of the Ho\v rava
theory requires to extend the phase space by adding the
canonical pairs $(N^{i}, \pi_{i})$ and the BFV ghost pairs $(C^{i},\bar{\mathcal{P}}_{i})$, $(\bar{C}_{i},\mathcal{P}^{i})$. The definition of the BFV path
integral, under a given gauge-fixing condition, is
\begin{eqnarray}
&&
Z =
\int \mathcal{D}V e^{iS} \,,
\label{Z}
\\
&&
\mathcal{D}V = 
\mathcal{D} g_{ij} \mathcal{D}\pi^{ij} \mathcal{D}N \mathcal{D} N^{k}\mathcal{D}\pi_{k}
\mathcal{D} C^i \mathcal{D} \bar{\mathcal{P}}_i
\mathcal{D} \bar{C}_i \mathcal{D} \mathcal{P}^i
\times \delta(\theta_{1}) \det\frac{ \delta \theta_1 }{ \delta N } \,, \;\;\;
\label{measure}
\\ &&
S=
\int dt d^{2}x \left( 
  \pi^{ij} \dot{g_{ij}} 
+ \pi_i \dot{N}^i 
+ \bar{\mathcal{P}}_{i} \dot{C}^{i} 
+ \mathcal{P}^{i} \dot{\bar{C}}_{i} 
- \mathcal{H}_{\Psi}
\right) \,.
\label{scan} 
\end{eqnarray}
The factor $\delta(\theta_1) \det(\delta\theta_1/\delta N)$ in (\ref{measure}) is the measure associated with the second-class constraints, see Ref.~\cite{Bellorin:2022efu}. Dirac delta $\delta(\theta_1)$ can be promoted to the quantum canonical Lagrangian by means of the integration on the Lagrange multiplier $\mathcal{A}$. The derivative $\delta\theta_1/\delta N$ can also be incorporated by means of a pair of fermionic ghosts, which we denote by $\eta,\bar{\eta}$. Thus, the measure of the second-class constraints becomes
\begin{equation}
\delta(\theta_{1}) \det\frac{ \delta \theta_1 }{ \delta N } =	
\int\mathcal{D}\mathcal{A} 
\mathcal{D}\bar{\eta} \mathcal{D}\eta
\exp\left[
i \int dt d^2x\left(
\mathcal{A}\theta_1
- \bar{\eta}  \frac{ \delta \theta_1 }{ \delta N } \eta
\right)
\right] \,.
\label{measurewithB}
\end{equation}
The path integral is defined by the measure and the quantum action:
\begin{eqnarray}
	&&
	\mathcal{D}V = 
	\mathcal{D} g_{ij} \mathcal{D}\pi^{ij} \mathcal{D}N \mathcal{D} N^{k}\mathcal{D}\pi_{k}
	\mathcal{D} C^i \mathcal{D} \bar{\mathcal{P}}_i
	\mathcal{D} \bar{C}_i \mathcal{D} \mathcal{P}^i
	\mathcal{D}\mathcal{A} 
	\mathcal{D}\bar{\eta} \mathcal{D}\eta \,, 
	\label{medida}
	\\ &&
	S=
	\int dt d^{2}x \left( 
	\pi^{ij} \dot{g_{ij}} 
	+ \pi_i \dot{N}^i 
	+ \bar{\mathcal{P}}_{i} \dot{C}^{i} 
	+ \mathcal{P}^{i} \dot{\bar{C}}_{i} 
	- \mathcal{H}_{\Psi}
	+ \mathcal{A}\theta_1
	- \bar{\eta}  \frac{ \delta \theta_1 }{ \delta N } \eta
	\right) \,.
	\label{quantumaction} 
\end{eqnarray}
The quantum gauge-fixed Hamiltonian density is defined by
\begin{equation}
 \mathcal{H}_{\Psi} =
 \mathcal{H}_0 
 + \{\Psi,\Omega\}_{\text{D}} \,,
 \label{bfvhamiltonian}
\end{equation}
where $\Omega$ is the generator of the BRST symmetry, given by
\begin{eqnarray}
    \Omega&=&
    \int\,d^{2}x
    \left(
    \mathcal{H}_kC^k
    +\pi_k\mathcal{P}^k
    -C^k\partial_kC^l\bar{\mathcal{P}}_l
    \right)\,.
\end{eqnarray}
$\Psi$ is a gauge-fixing fermionic function and $\{\,,\}_{\text{D}}$ indicates Dirac brackets.


\subsection{Gauge fixing and propagators}
The original aim in the BFV formalism was to incorporate covariant gauges to the process of unitary quantization in relativistic theories. These covariant gauges are noncanonical in the basic phase space. The BFV quantization was developed to cover these cases. In the case of the Ho\v{r}ava theory, it turns out that the same functional structure of the gauge used in the BFV quantization is appropriate \cite{Bellorin:2021tkk,Bellorin:2021udn}. The BFV gauge-fixing condition and its associated fermionic function $\Psi$ are of the general form
\begin{eqnarray}
 &&
 \dot{N}^i - \chi^i = 0\,,
 \label{relativisticgaugephi}
 \\ &&
 \Psi =
 \bar{\mathcal{P}}_{i} N^{i} + \bar{C}_{i} \chi^{i} \,.
 \label{relativisticgaugepsi}
\end{eqnarray} 
We introduce perturbative variables by making perturbations around a flat background: $g_{ij} = \delta_{ij}$, $N =1$, and the rest of variables taking zero value. The perturbations are denoted by
\begin{equation}
 g_{ij} = \delta_{ij} + h_{ij} \,,
 \quad
 N = 1 + n \,,
 \quad
 N^i = n^i \,,
\end{equation}
and for the rest of perturbative field variables we keep the original notation. For the $2+1$ Ho\v{r}ava theory we choose $\chi^{i}$ to be
\begin{eqnarray}
&&
\chi^{i} = \rho\mathfrak{D}^{ij}\pi_{j} - 2\rho\left(\Delta\partial_{j}h_{ij} - \lambda\bar{\kappa}\Delta\partial_{i}h
+ \kappa\partial_{ijk}h_{jk}\right),
\label{gaugefixing}
\\ &&
\mathfrak{D}^{ij} \equiv 
\delta_{ij} \Delta + \kappa \partial_{ij} \,.
\end{eqnarray}
with $\bar{\kappa} = \kappa + 1$. The coefficients in the gauge fixing are chosen to simplify and ensure regularity of the resulting propagators. In the following we use more combinations of constants,
\begin{equation}
\rho_1 = 2 (1-\lambda)\bar{\kappa} \,,
\quad
\rho_2 = 4 \alpha_{1}-\frac{\alpha_{5}^{2}}{\alpha_{67}} \,.
\end{equation}
The gauge-fixed Hamiltonian is
\begin{eqnarray}
    \mathcal{H}_{\Psi} &=&
 \mathcal{H}_{0} 
+ \mathcal{H}_in^{i}
+ \bar{\mathcal{P}}_i \mathcal{P}^{i}
- \bar{\mathcal{P}}_{i} ( n^{j} \partial_{j} C^{i} 
+ n^{i}\partial_{j} C^{j} )
+ \rho\pi_i \mathfrak{D}^{ij} \pi_{j}
\nonumber\\
&&
- 2 \rho\pi_i ( 
\Delta\partial_jh_{ij}
- \lambda \bar{\kappa} \Delta\partial_{i}h
+ \kappa \partial_{ijk} h_{jk} )
+ \bar{C}_i\{\chi^{i} \,, \mathcal{H}_j\} C^{j}.
\end{eqnarray}

With the gauge condition (\ref{gaugefixing}), most of the propagators get a regular form. The definition of regular propagators was introduced in Ref.~\cite{Anselmi:2008bq} to study the renormalizability of nonrelativistic gauge theories. This was used in the proof of renormalization of the projectable Ho\v{r}ava theory \cite{Barvinsky:2015kil,Barvinsky:2017zlx}. In Euclidean space-time, consider a propagator between two fields that have scaling dimensions $r_1$ and $r_2$. It is regular if it is given by the sum of terms of the form
\begin{equation}
	\frac{ P(\omega,k^i) }{ D(\omega,k^i) } \,,
 \label{defregular}
\end{equation}
where $D$ is the product
\begin{equation}
	D = 
	\prod_{m=1}^{M} ( A_m \omega^2 + B_m k^{2d} + \cdots ) \,, 
	\quad
	A_m \,, B_m > 0 \,,
	\label{regular}
\end{equation}
and $P(\omega,k^i)$ is a polynomial of maximal scaling degree less than or equal to $r_1 + r_2 + 2(M-1)d$.

We present all the nonzero propagators. In Fourier space, after a Wick rotation, they are
\begin{equation}
\begin{split}
\langle \bar{\mathcal{P}}_i\mathcal{P}^{j}\rangle =&		
4 \rho k^4 \left( P_{ij}\mathcal{T}_{3} 
+ 2 \rho_1 \hat{k}_{i} \hat{k}_{j} \mathcal{T}_{2} \right)
\,,
\\
\langle \mathcal{P}_i\bar{C}^{j}\rangle =& 
- \langle \bar{\mathcal{P}}_iC^{j}\rangle =
\omega \mathcal{S}_{ij} \,,
\quad
\langle \bar{C}_iC^{j}\rangle = - \mathcal{S}_{ij}  \,,
\\
\langle p^{ij}p^{kl} \rangle =&
\rho_2 k^{4} P_{ij} P_{kl}\mathcal{T}_{1} \,,
\\
\langle h_{ij} h_{kl} \rangle =&
8 \left(\omega^2 - 2\rho k^{4}\right) Q_{ijkl}\mathcal{T}^{2}_{3}
+ 4 \left( {\sigma} P_{ij} P_{kl}
+ \bar{\sigma} \hat{k}_i \hat{k}_j P_{kl} 
+ \bar{\sigma} \hat{k}_k \hat{k}_l P_{ij} 
\right) \mathcal{T}_{1}
+ 4 \sigma Q \hat{k}_{i} \hat{k}_{j} \hat{k}_{k} \hat{k}_{l} \mathcal{T}_{1} \mathcal{T}_{2}^{2} \,,
\\
\langle h_{ij} p^{kl} \rangle =&
2\omega \left[ 2Q_{ijkl} \mathcal{T}_{3}
+ P_{ij} P_{kl}\mathcal{T}_{1}
+ 4 \rho \bar{\sigma}\bar{\kappa} \left(1- 2\lambda +  \rho_2 \right) k^2 k_{i}k_{j} P_{kl} \mathcal{T}_{1}\mathcal{T}_{2}
+  \hat{k}_{i} \hat{k}_{j} \hat{k}_{k} \hat{k}_{l} \mathcal{T}_{2} \right] \,,
\\
\langle n^{i} h_{jk} \rangle =&
16 i \omega \rho k^3 \left( P_{i(j} \hat{k}_{k)} \mathcal{T}_{3}^{2}
+ \bar{\kappa} \hat{k}_i \hat{k}_j \hat{k}_k \mathcal{T}_{2}^{2} \right) \,,
\\
\langle n^{i} p^{jk} \rangle =&
2 i k^3 \left[ 2 P_{i(j} \hat{k}_{k)} \mathcal{T}_{3}
+ \hat{k}_{i} \left( \rho_1 \hat{k}_j \hat{k}_{k}
- 2 \lambda\bar{\kappa} P_{jk} \right) \mathcal{T}_{2} 
\right] \,,
\\
\langle n^{i} n^{j} \rangle =&
- 4 \rho k^2 P_{ij} \left(\omega^{2}-2\rho k^4\right)\mathcal{T}^{2}_{3}
- 4 \rho \bar{\kappa} \left(\omega^{2} - 2\rho\rho_{1}k^{4} \right) k_{i}k_{j} \mathcal{T}_{2}^{2} \,,
\\ 
\langle h_{ij} n \rangle =&
\frac{2\alpha_{5}}{\alpha_{67}} \left( \sigma P_{ij}
+ \bar{\sigma} \hat{k}_i \hat{k}_j \right)\mathcal{T}_{1} \,,
\quad
\langle p^{ij} n \rangle =
\frac{\alpha_{5}}{\alpha_{67}}\omega P_{ij}\mathcal{T}_{1} \,,
\quad
\langle n^{i}\pi_{j} \rangle =
\omega \mathcal{S}_{ij}  \,,
\\
\langle \pi_i h_{jk} \rangle =&
- 4 i \left( P_{i(j} k_{k)} \mathcal{T}_{3}
+ k_{i} \hat{k}_{j} \hat{k}_{k}\mathcal{T}_{2} \right) \,,
\quad
\langle nn \rangle =
\frac{ \alpha_{5}^2 \sigma }{ \alpha_{67}^2 } \mathcal{T}_{1} \,,
\end{split}
\label{propagators}
\end{equation}
and 
\begin{equation}
 \langle \mathcal{A}\mathcal{A} \rangle =
 \langle \mathcal{A} n \rangle = 
 \langle \bar{\eta} \eta \rangle = 
 \frac{1}{ \alpha_{67} k^{4}} \,,
 \label{irregularprop}
\end{equation}
where
\begin{equation}
\begin{split}
&
\hat{k}_i = \frac{ k_i }{k} \,,
\quad
P_{ij} = \delta_{ij} - \hat{k}_{i} \hat{k}_{j} \,,
\\ &
Q_{ijkl} =
\hat{k}_i \hat{k}_k P_{jl} + \hat{k}_j \hat{k}_k P_{il} 
+ \hat{k}_i \hat{k}_l P_{jk} + \hat{k}_j \hat{k}_l P_{ik} \,,
\\ &
Q =
\omega^{4} 
+ \left[ 4\rho\left(2\lambda^2 +2\lambda -1 \right)\bar{\kappa}
- \rho_2 \right] 
\frac{\omega^{2}k^{4}}{1-\lambda} 
+ 4 \bar{\kappa} \rho \left( \rho_2 + 4\lambda^{2}\bar{\kappa}\rho \right) k^8  \,,
\\ &
\mathcal{T}_{1} =
\left(\omega^2-\rho_2 {\sigma}k^4\right)^{-1} \,,
\quad
\mathcal{T}_{2} = 
\left(\omega^{2} + 2 \rho\rho_1 k^4 \right)^{-1},
\\ &
\mathcal{T}_{3} =
\left(\omega^{2} + 2\rho k^{4}\right)^{-1},
\quad
\mathcal{S}_{ij} =
2 P_{ij}\mathcal{T}_{3} + 2 \hat{k}_{i} \hat{k}_{j} \mathcal{T}_{2} \,.
\end{split}
\end{equation}
All the propagators in the list (\ref{propagators}) satisfy the condition of regularity.\footnote{Actually, the condition of regular propagator in (\ref{regular}) requires the coefficients in the denominators to be strictly positive. In the present theory this is satisfied if $\rho > 0$, $\rho_1 > 0$ and $\sigma \rho_2 < 0$.} On the contrary, the three propagators $\langle \mathcal{A} \mathcal{A} \rangle$,  $\langle \mathcal{A} n \rangle$, and $\langle \bar{\eta}\eta \rangle$ in (\ref{irregularprop}), which are the only ones that involve the fields  $\mathcal{A},\eta,\bar{\eta}$ associated with the measure of the second-class constraints, are independent of $\omega$; hence the propagators in (\ref{irregularprop}) are irregular.


\subsection{The $(3+1)$-dimensional case}
In this section we discuss about the extension of the analysis we have done to the case of $3$ spatial dimensions. The BFV quantization in the $(3+1)$-dimensional case, with the appropriate gauge-fixing condition, has been done in Ref.~\cite{Bellorin:2021udn}. Since our interest is in the consistent quantization and the characterization of the regular propagators, we may take the potential of theory only with the terms that contribute to the propagators, and, among them, only with the terms of $z=3$ order. These terms are the ones on which the definition of regular propagator is based on, besides the time-derivative terms (see Ref.~\cite{Barvinsky:2015kil}). The $z=3$ order terms are the ones that dominate at the ultraviolet regime. Therefore, we take the potential \cite{Colombo:2014lta}
\begin{equation}
\mathcal{V}=
-\alpha_{3}\nabla^{2}R\nabla_{i}a^{i}-\alpha_{4}\nabla^{2}a_{i}\nabla^{2}a^{i}-\beta_{3}\nabla_{i}R_{jk}\nabla^{i}R^{jk}-\beta_{4}\nabla_{i}R\nabla^{i}R \,,
\end{equation}
where $\alpha_3,\alpha_4,\beta_3,\beta_4$ are coupling constants. The definition of $\mathcal{V}$ determines the form of the Hamiltonian and the constraint $\theta_1$ (see \cite{Bellorin:2021udn}). The constraint takes the form
\begin{equation}
\theta_1 \equiv \frac{N}{\sqrt{g}}\left(\pi^{ij}\pi_{ij}
+\frac{\lambda}{1-3\lambda}\pi^{2}\right)
+\sqrt{g} N \mathcal{V}	
-\alpha_{3}\sqrt{g} \nabla^{2}(N\nabla^{2}R)
+2\alpha_{4} \sqrt{g} \left(\nabla^{i}\nabla^{2}(N\nabla^{2}a_{i})\right)\,.
\label{hamiltonianconst}
\end{equation}
The dimensionality of the foliation requires a specific degree of anisotropy for the gauge-fixing condition. In this case, the gauge-fixing condition is given by
\begin{equation}\label{gauge3+1}
\chi^{i}= 
\rho\mathfrak{D}^{ij}\pi_{j}
- 2 \rho \Delta^{2}\partial_{j}h_{ij}
+ 2 \rho \lambda(1+\kappa) \Delta^{2}\partial_{i}h
- 2 \kappa \rho \Delta\partial_{i}\partial_{j}\partial_{k}h_{jk} 
\,.
\end{equation}
where
\begin{equation}
\mathfrak{D}^{ij} = \delta_{ij} \Delta^2 
+ \kappa \Delta \partial_i \partial_j 
\,.
\end{equation}

The BFV quantization of the $3+1$ theory was brought to a Lagrangian form in Ref.~\cite{Bellorin:2021udn}, by integrating over all canonical momenta. The resulting Lagrangian is defined on the quantum fields $h_{ij}$, $n$, $n_i$, the BFV ghosts $C^i,\bar{C}_i$, and the variables associated with the second class constraints $\mathcal{A}$, and $\eta,\bar{\eta}$. To present the propagators, we perform the transverse-longitudinal decomposition
\begin{equation}\label{decomposition}
h_{ij} = h_{ij}^{TT}
+\frac{1}{2}\left(\delta_{ij} - \frac{\partial_i\partial_j}{\Delta}\right)h^{T}
+ \partial_{(i}h_{j)} \,,
\end{equation}
and we also decompose the vectors, 
\begin{equation}
\begin{split}
& n^{i}= n^{i}{}^{T} + \partial^{i}n^L \,, 
\qquad \partial_in^{i}{}^{T}=0 \,,
\\ &
h_{i}=
h_{i}{}^{L}+\partial_{i}\Delta^{-1}h^L \,, 
\qquad \partial_i h_{i}{}^{L}=0 \,,
\end{split}
\label{vectordecomposition}
\end{equation}
and similarly with the vectors $C^i,\bar{C}_i$. The resulting (nonzero) propagators are 
\begin{equation}
\begin{split}
&
 \langle h_{ij}^{TT}h_{ij}^{TT}\rangle = -4\mathcal{P}_{1},\quad k^{2}\langle h_{i}^{L}h_{i}^{L}\rangle = -8\mathcal{P}_{3},\quad \langle h^{T}h^{T}\rangle = -8\frac{1-\lambda}{1-3\lambda}\mathcal{P}_{2},
\\  &
\langle h^{T}h^{L}\rangle = -\frac{8\lambda}{1-3\lambda}\mathcal{P}_{2},\quad \langle h^{L}h^{L}\rangle = -\frac{4}{1-\lambda}\mathcal{P}_{4} - \frac{8\lambda^{2}}{(1-\lambda)(1-3\lambda)}\mathcal{P}_{2},
\\ &
\langle n h^{T}\rangle = 4\frac{\alpha_3}{\alpha_4}\frac{(1-\lambda)}{(1-3\lambda)}\mathcal{P}_2\,,\quad \langle n h^{L}\rangle = 4\frac{\alpha_{3}}{\alpha_{4}}\frac{\lambda}{(1-3\lambda)}\mathcal{P}_2,\quad
\langle nn\rangle = - 2 \frac{(1-\lambda)}{(1-3\lambda)}\left(\frac{\alpha_3}{\alpha_4}\right)^2\mathcal{P}_2\,,
\\ &
\langle n_{i}^{T}n_{i}^{T}\rangle = - 4\rho k^4 \mathcal{P}_3,\quad  \langle n^{L}n^{L}\rangle = - 4\rho\bar{\kappa}k^4\mathcal{P}_4\,,\quad \langle \bar{C}_{k}^{T}C^{kT}\rangle = \mathcal{P}_3\,,\quad k^{2}\langle \bar{C}^{L}C^{L}\rangle = \mathcal{P}_4 \,,
\end{split}
\label{propagtaors3+1}
\end{equation}
and
\begin{equation}
\langle \mathcal{A}\mathcal{A}\rangle = \langle n\mathcal{A}\rangle= \langle \bar{\eta}\eta\rangle=-\mathcal{P}_5
\label{irregular3+1}
\,,
\end{equation}
where 
\begin{equation}
\begin{split}
\mathcal{P}_1 =& \frac{1}{\omega^{2} - \beta_{3}k^{6}} \,, \quad
\mathcal{P}_2 = \left[ \omega^{2}
- \frac{(1-\lambda)}{\alpha_{4}(1-3\lambda)}
\left(\alpha_{4} \left( 3\beta_{3} + 8 \beta_{4} \right) - 2 \alpha_{3}^{2}\right) k^{6} \right]^{-1} \,,
\\ 
\mathcal{P}_3 =& \frac{1}{\omega^{2} - 2\rho k^{6}} \,,
\quad
\mathcal{P}_4 = \frac{1}{\omega^{2} - 4\bar{\kappa}(1-\lambda)k^{6}} \,,\quad
\mathcal{P}_5 = \frac{1}{\alpha_{4}k^{6}} \,.
\end{split}
\label{Pfactors}
\end{equation}
In the above list, all propagators are proportional to the factors given in (\ref{Pfactors}). Among these factors, only $\mathcal{P}_5$ has an unavoidable irregular form, as happened in the $(2+1)$ case. Actually, the condition of regularity on the factors $\mathcal{P}_1$ to $\mathcal{P}_4$ requires some restrictions on the coupling constants. Specifically, condition (\ref{regular}) requires that the coefficients of the $k^6$ order terms in the denominators must be strictly positive. Hence, the only irregular propagators are the ones of the fields associated with the second-class constraints, given in (\ref{irregular3+1}). In this sense, the situation is analogous to the $(2+1)$ case.


\section{The effective action}
\subsection{The role of the second-class constraints}
We obtain the one-loop effective action corresponding to the path integral defined in (\ref{medida}) -- (\ref{quantumaction}). Actually, the formalism can be adapted directly to any dimension, once the potential that defines the Lagrangian is given. We denote all quantum fields, including ghosts, by $\Phi^\mathcal{I}$. We introduce the effective action $\Gamma[\hat{\Phi}^\mathcal{I}]$, where $\hat{\Phi}^\mathcal{I}$ denotes the effective fields, in the usual way: if $\mathcal{J}_{\mathcal{I}}$ represents a source for each field, then
\begin{eqnarray}
    Z[\mathcal{J}_{\mathcal{I}}] =
    \exp\left( i W[\mathcal{J}_{\mathcal{I}}]\right)
    &=&
    \int\mathcal{D}\Phi^{\mathcal{I}} \exp\left( iS[\Phi^{\mathcal{I}}]
    + i \int dtd^{2}x\Phi^{\mathcal{I}}\mathcal{J}_{\mathcal{I}}\right)
    \\
    &=&
    \exp\left[ i \left(\Gamma[\hat{\Phi}^{\mathcal{I}}]
    + \int dtd^{2}x\hat{\Phi}^{\mathcal{I}}\mathcal{J}_{\mathcal{I}}\right)\right]
    \,,
\end{eqnarray}
where 
\begin{equation}
 \Gamma[\hat{\Phi}^{\mathcal{I}}] = 
 W - \int dtd^{2}x\, \hat{\Phi}^{\mathcal{I}}\mathcal{J}_{\mathcal{I}} \,.
\end{equation}
The one-loop quantum corrections are contained in the effective action $ \bar{\Gamma}^{(1)}[\hat{\Phi}] = \Gamma^{(1)}[\hat{\Phi}] - S[\hat{\Phi}]$. We denote derivatives of $S[\hat{\Phi}]$ by
\begin{equation}
 S_{\mathcal{I}\cdots\mathcal{J}} = 
 \frac{ \partial S }
 { \partial\hat{\Phi}^{\mathcal{I}} \cdots \partial\hat{\Phi}^{\mathcal{J}} } \,.
\end{equation}
We use a convention for left/right derivatives in Grassmann variables. Derivatives with respect to ghosts with a bar ($\bar{C}_i$, $\bar{\eta}$) are left derivatives, and derivatives with respect to ghosts without bar ($C^i,\eta$) are right derivatives. $\bar{\Gamma}^{(1)}[\hat{\Phi}]$ is given by the formula
\begin{equation}
\exp\left(
i\bar{\Gamma}^{(1)}
\right)
=
\int\mathcal{D}\Phi^{\mathcal{I}} \exp \left( \frac{i}{2}   S_{\mathcal{I}\mathcal{J}} \Phi^{\mathcal{J}} \Phi^{\mathcal{I}}
\right) \,.
\label{gammabar}
\end{equation}
For each contraction of the field indices $\mathcal{I},\mathcal{J}$ there is an integration on time and space that we have not written.

To evaluate (\ref{gammabar}), we separate ghost from nonghost fields. Let us introduce an index notation to this end. Indices like $I,J = \{ g_{ij},\pi^{ij},N,N^i,\pi_i\}$ run for nonghost (bosonic) fields and indices like $\alpha,\beta = \{ C^i,\bar{\mathcal{P}}_i, \bar{C}_i, \mathcal{P}^i,\eta,\bar{\eta} \}$ run for ghost fields (the same for quantum and effective fields). We perform on (\ref{gammabar}) a procedure of expanding in ghost fields, completing squares on them, and integrating on all fields:
\begin{eqnarray}
\exp\left(
    i\bar{\Gamma}^{(1)}
     \right)
     &=&
     \int\mathcal{D}\Phi^{I}\mathcal{D}\Phi^{\alpha}\mathcal{D}\Phi^{\beta}
     \exp
     \left[
     - i\left( \Phi^{\alpha} 
     - S_{I\sigma} \Phi^{I} S_{\sigma\alpha}^{-1} \right)
     S_{\alpha\beta}
     \left( \Phi^{\beta} - S_{\beta\rho}^{-1} S_{\rho J} \Phi^{J} \right)
     \right.
     \nonumber\\
     &&
     \left.
     + \frac{i}{2} S_{IJ}\Phi^{J}\Phi^{I}
     + i S_{I\alpha} S_{\alpha\beta}^{-1} S_{\beta J}\Phi^{J}\Phi^{I}\right]
     \nonumber\\
     &=&
     \det\left(S_{\alpha\beta}\right)
     \left[\det
     \left(S_{IJ} 
     + 2 S_{I\alpha}S_{\alpha\beta}^{-1}S_{\beta J} 
     \right)
     \right]^{-1/2}\,.
     \label{effectiveactionpre}
\end{eqnarray}
We define the matrices
\begin{equation}
\mathbb{A}_{IJ} = S_{IJ} \,,
\quad
\mathbb{B}_{I \alpha} = S_{I \alpha} \,,
\quad
\mathbb{D}_{\alpha\beta} = S_{\alpha\beta} \,,
\end{equation}
such that the effective action (\ref{effectiveactionpre}) is given in terms of the Berezinian,
\begin{equation}
i\bar{\Gamma}^{(1)} =
- \frac{1}{2} \ln \left( 
\frac{ \det ( \mathbb{A} - 2 \mathbb{B} \mathbb{D}^{-1} \mathbb{B}^T ) }
{ \det \mathbb{D} } \right)  \,.
\label{Berfull}
\end{equation}
The factor $\det\mathbb{D}$ in the denominator contains derivatives exclusively with respect to ghost fields. Among them, derivatives with respect to $\eta,\bar{\eta}$ play a central role in the cancellation of irregular loops. For this purpose, we separate the two types of ghost fields, owing on the fact that there is no derivative entry $S_{\alpha\beta}$ mixing BFV ghosts with the $\eta,\bar{\eta}$ ghosts, since there is no coupling between these fields. Let us introduce hatted indices $\hat{\alpha},\hat{\beta} = \{ C^i,\bar{\mathcal{P}}_i, \bar{C}_i, \mathcal{P}^i \}$ exclusively for the BFV ghosts. We define the matrices
\begin{equation}
\begin{split}
 & \mathbb{H}_{\hat{\alpha}\hat{\beta}} = S_{\hat{\alpha}\hat{\beta}} \,,
   \quad
   \mathbb{E}_{I\hat{\alpha}} = S_{I\hat{\alpha}} \,,
 \\
 & \mathbb{F}_{\alpha\beta} = S_{\alpha\beta} \,,
   \quad
   \mathbb{C}_{I\alpha} = S_{I\alpha} 
 \quad \mbox{only for} \quad \alpha,\beta = \eta,\bar{\eta} \,. 
\end{split} 
\end{equation}
Thus, the effective action can be written in the expanded form
\begin{equation}
 i\bar{\Gamma}^{(1)} =
 - \frac{1}{2} \ln \left( 
 \frac{ \det ( \mathbb{A} - 2 \mathbb{E} \mathbb{H}^{-1} \mathbb{E}^T
 - 2 \mathbb{C} \mathbb{F}^{-1} \mathbb{C}^T )}
 { \det \mathbb{F} \det \mathbb{H} } \right)  \,.
 \label{generator}
\end{equation}

We remark that the result for the effective action, (\ref{Berfull}) or (\ref{generator}), takes the same form in any spatial dimension, since the formalism to obtain the effective action is general. 


\subsection{Cancellations of divergences}
\subsubsection{Regular versus irregular loops}
In the list of propagators shown in (\ref{propagators})--(\ref{irregularprop}), the only irregular ones are $\left< \mathcal{A} \mathcal{A} \right>$, $\left< \mathcal{A} n \right>$ and $\left< \bar{\eta} \eta \right>$. It is remarkable that they involve field variables associated with the second-class constraints. The three propagators are equal to $1/\alpha_{67} k^4$, which is a radically irregular form since it does not depend on the frequency $\omega$. Fortunately, the irregularity of these propagators is automatically cured if, at least, one regular propagator enters in the loop. Consider a loop formed by an arbitrary number $r$ of irregular propagators and one regular propagator. The loop integral is proportional to
\begin{equation}
\int d\omega\, d^2 k \frac{1}{ (k^4)^r } \frac{P(\omega,k^i)}{D(\omega,k^i)} V(k^i) \,,
\end{equation}
where $P(\omega,k^i)$ and $D(\omega,k^i)$ refer to the regular structure of propagators defined in (\ref{defregular})--(\ref{regular}), and $V(k^i)$ is the contribution of the vertices.\footnote{This is a quantization in the Hamiltonian formalism, hence vertices do not depend on $\omega$.} The factor $P(\omega,k^i)/D(\omega,k^i)$ coming from the regular propagator is sufficient to give a regular structure to the integration on $\omega$, regardless of the fact that none of the irregular propagators depend on it. The integration on $\omega$ results the same as if the loop had only the regular propagator; hence the issue of the divergence in $\omega$ is the same as the case of only regular propagators, the irregular propagators do not alter this. More regular propagators in the loop preserve the regular structure of the integration on $\omega$ and $k^i$.

On the other hand, dangerous divergences arise if the loop is made exclusively of irregular propagators, as we discussed in Ref.~\cite{Bellorin:2022qeu}. We call irregular loop to this case, and regular loop when at least one regular propagator is present. In the case of an irregular loop, the integral is proportional to
\begin{equation}
\int d\omega d^2 k \frac{1}{ (k^4)^r } V(k^i)
= \Lambda_\omega \int d^2 k \frac{1}{ (k^4)^r } V(k^i) \,.
\end{equation}
The divergence $\Lambda_\omega$ multiplies the whole diagram, hence it cannot be renormalized by counterterms. This is an unavoidable result of any irregular loop that, on the other hand, it is not exhibited by regular loops.  Therefore, to save the theory it is necessary that irregular loops be canceled completely.

We show the exact cancellation of irregular loops in the framework of the effective action. In the next subsection we give preliminary evaluations of the first and second derivative of the effective action to illustrate how the cancellations occur. We give the general proof for arbitrary order in derivatives at the end of this section. The analysis holds in any spatial dimension of the foliation, once the identification of the irregular propagators has been done. Indeed, in the following we do not require the explicit form of the propagators, only the relations between them. We have shown that the $(3+1)$ case is analogous to the $(2+1)$ case, since the irregular propagators are associated with the same quantum fields of the second-class constraints. The only difference arising in the computation of the cancellations of divergences is that the range of the indices associated with the spatial dimensionality must be augmented, for example in $h_{ij}$, $n_i$, the ghost $C^i$, and so on. Other indices that include the spatial ones, as the ones arising in the next section, must be adapted as well.

\subsubsection{First and second derivatives of the effective action}
To establish a relation with Feynman diagrams, we evaluate derivatives of the effective action at zero effective fields. Propagators are contained in entries of inverse matrices like $\mathbb{A}^{-1}_{IJ}|_0$, and vertices are higher order derivatives $S_{\mathcal{I}\mathcal{J}\mathcal{K}\cdots}|_0$.\footnote{Entries of inverse matrices are equal to $\mathbb{A}^{-1}_{IJ}|_0 = 2 \left< \Phi^I \Phi^J \right>$. Despite this factor $2$, in the whole discussion we call $\mathbb{A}^{-1}_{IJ}|_0$ directly the propagator (we do not make the substitution $\mathbb{A}^{-1}_{IJ}|_0 = 2 \left< \Phi^I \Phi^J \right>$ explicitly at any stage).} There are various identities due to the structure of the Lagrangian that we require \cite{Bellorin:2022qeu}. First, $\det \mathbb{A}$ satisfies 
\begin{equation}
	\left[ \det \mathbb{A} = 
	- S_{nn} \mbox{Minor}(\mathbb{A})_{\mathcal{A} \mathcal{A}} = 
	- S_{nn} \mbox{Minor}(\mathbb{A})_{\mathcal{A} n} \right]_{0} \,.
	\label{identity1}
\end{equation}
Let us introduce hatted bosonic indices $\hat{I},\hat{J}$, which run for the same values of $I,J$, excluding the field $\mathcal{A}$. Then, we have the identity
\begin{equation}
	S_{\mathcal{A} \hat{I}}|_0 = - S_{n \hat{I}}|_{0} \,.
	\label{identity2}
\end{equation} 
The third identity we need is
\begin{equation}
	S_{\mathcal{A}n} = S_{\bar{\eta}{\eta}} 
	= \frac{ \delta \theta }{ \delta n} \,,
	\label{equalirregular}
\end{equation}
which holds at all orders, and it is a direct consequence of the fact that the dependence of the action in $\mathcal{A}$ and $\eta,\bar{\eta}$ comes only from the terms in Eq.~(\ref{measurewithB}). Due to the block-diagonal structure of $S_{\mathcal{I}\mathcal{J}}|_0$, the factor $1/S_{\bar{\eta}{\eta}}|_0$ is directly the irregular propagator $\left< \bar{\eta} \eta \right>$. Similarly, from relations (\ref{identity1}) and (\ref{identity2}), we obtain
\begin{equation}
\left. \mathbb{A}^{-1}_{\mathcal{A}n} \right|_0 =
\frac{1}{ \left. S_{\mathcal{A} n} \right|_0}  \,,
\label{ProAn}
\end{equation}
hence $1/S_{\mathcal{A} n} |_0$ is the irregular propagator $\left< \mathcal{A} n \right>$.

The first derivative of the effective action (\ref{generator}) with respect to an arbitrary field $\Phi^{\mathcal{I}}$ results
\begin{equation}
\begin{split}
	-2i\frac{ \delta \bar{\Gamma}^{(1)} }{ \delta\Phi^{\mathcal{I}} } 
    =&
	\left( \mathbb{A}' - 2 \mathbb{C} \mathbb{F}^{-1} \mathbb{C}^T \right)^{-1}_{JK} 
	\left( \frac{ \delta }{	\delta\Phi^{\mathcal{I}} } \mathbb{A}_{KJ}'  
	- 2 \frac{ \delta }{	\delta\Phi^{\mathcal{I}} } \left( \mathbb{C} \mathbb{F}^{-1} \mathbb{C}^T \right)_{KJ}
	\right)
 \\
 &
    - \mathbb{H}^{-1}_{\hat{\alpha}\hat{\beta}} \frac{\delta}{\delta\Phi^{\mathcal{I}}} \mathbb{H}_{\hat{\beta}\hat{\alpha}}
	- 2 \frac{ S_{\mathcal{I} \bar{\eta} \eta } }{ S_{\bar{\eta} \eta} } \,,
\end{split}
\end{equation}
where $\mathbb{A}' =
\mathbb{A} - 2 \mathbb{E} \mathbb{H}^{-1} \mathbb{E}^T$, and we have used $\det \mathbb{F} = 
\left( S_{\bar{\eta}\eta} \right)^2$. Note that the last term, when evaluated at zero, is an irregular propagator with the corresponding vertex. The mixed matrices $\mathbb{C}$ and $\mathbb{E}$ have no zeroth-order term; they start at linear order. Hence
\begin{eqnarray}
	\mathbb{C}|_0 = 0 \,,
	\quad  
	\left. \frac{ \delta }{ \delta\Phi^{\mathcal{I}} } \left( \mathbb{C} \mathbb{F}^{-1} \mathbb{C}^T \right) \right|_0 = 0 \,,
\\	
	\mathbb{E}|_0 =  0 \,,
	\quad 
	\left. \frac{ \delta }{ \delta\Phi^{\mathcal{I}} } \left( \mathbb{E} \mathbb{H}^{-1} \mathbb{E}^T \right) \right|_0 = 0 \,.
\end{eqnarray}
Thanks to this, the derivative of the effective action simplifies to
	\begin{eqnarray}
	\left. -2i\frac{ \delta \bar{\Gamma}^{(1)} }{ \delta\Phi^{\mathcal{I}} } \right|_0 
 &=&
    \left[
	\mathbb{A}^{-1}_{JK} S_{\mathcal{I}KJ} 
    - \mathbb{H}^{-1}_{\hat{\alpha}\hat{\beta}} 
	S_{\mathcal{I}\hat{\beta}\hat{\alpha}}
	- 2 \frac{ S_{\mathcal{I} \bar{\eta} \eta } }{ S_{\bar{\eta} \eta} } 
	\right]_0
	\nonumber
	\\
	&=&
	\left[
	\mathbb{A}^{-1}_{(JK)^*} S_{\mathcal{I} (KJ)^*}
	- \mathbb{H}^{-1}_{\hat{\alpha}\hat{\beta}} 
	S_{\mathcal{I}\hat{\beta}\hat{\alpha}}
	+ 2 \frac{ S_{\mathcal{I} \mathcal{A}n} }{ S_{\mathcal{A} n} }
	- 2 \frac{ S_{\mathcal{I} \bar{\eta} \eta } }{ S_{\bar{\eta} \eta} } 
	\right]_0 	\,,
	\label{dgenerator2}
	\end{eqnarray}
where $(JK)^*$ means that in the summation the component $\mathcal{A}n$ is excluded, and we have used (\ref{ProAn}). At this point we can see the cancellation between irregular loops. Identity (\ref{equalirregular}) implies $S_{\mathcal{A} n} = S_{\bar{\eta} \eta}$ and $ S_{\mathcal{I} \mathcal{A}n} =  S_{\mathcal{I} \bar{\eta} \eta}$ at all orders, hence the last two terms in (\ref{dgenerator2}) are exactly equal and cancel between them. The resulting derivative is
\begin{equation}
	\left. -2i\frac{ \delta \bar{\Gamma}^{(1)} }{ \delta\Phi^{\mathcal{I}} } \right|_0 =
	\left[
	\mathbb{A}^{-1}_{(JK)^*} S_{\mathcal{I} (KJ)^*}
    - \mathbb{H}^{-1}_{\hat{\alpha}\hat{\beta}} 
	S_{\mathcal{I}\hat{\beta}\hat{\alpha}}
	\right]_0 \,.
	\label{finalderivative}
\end{equation}
After the cancellation we have shown, expression (\ref{finalderivative}) contains exclusively regular propagators. The entries of the inverse matrix $\mathbb{A}^{-1}_{IJ}|_0$ are the propagators of the bosonic fields, and the entries $\mathbb{H}^{-1}_{\hat{\alpha}\hat{\beta}}|_0$ are the propagators of the BFV ghosts. Factors $S_{\mathcal{I} (KJ)^*}|_0$ and $	S_{\mathcal{I}\hat{\beta}\hat{\alpha}}|_0$ are three-leg vertices.  In the BFV-ghost sector all propagators are regular. The only irregular propagators in the bosonic sector are $\left< \mathcal{A} n \right> = \mathbb{A}^{-1}_{\mathcal{A} n }|_0 $ and $\left<\mathcal{A} \mathcal{A} \right> = \mathbb{A}^{-1}_{\mathcal{A} \mathcal{A} }|_0 $. The first one has been excluded in the expansion in $(JK)^*$, and the last one does not arise due to the fact that the action is of linear order in $\mathcal{A}$, hence $S_{\mathcal{I} \mathcal{A} \mathcal{A}} = 0$. 

We have seen that in the case of the first derivative no regular propagators remain at all. We now compute the second derivative of the effective action (\ref{generator}). This case is illustrative since irregular propagators do remain, but they cannot form a complete irregular loop, since they are necessary accompanied by one regular propagator to form a loop. To show this we require an additional identity restricting the possible vertices:
\begin{equation}
 S_{\mathcal{I}\mathcal{A}\alpha} = 0 \,,
 \label{novertex}
\end{equation} 
which is due to the fact that there is no coupling between the field $\mathcal{A}$ and the ghosts $\eta,\bar{\eta}$ or the BFV ghosts.

When evaluated at zero effective fields, the second derivative results
\begin{equation}
 \begin{split}
 \left.-2i\frac{\delta \bar{\Gamma}^{(1)}}{\delta\Phi^{\mathcal{J}}\delta\Phi^{\mathcal{I}}}\right|_{0}
 =&
 - \mathbb{A}^{-1}_{JM} S_{\mathcal{J}MN} 
   \mathbb{A}^{-1}_{NK} S_{\mathcal{I}KJ}
 + \mathbb{A}^{-1}_{JK} S_{\mathcal{J}\mathcal{I}KJ}
 - 4\mathbb{A}^{-1}_{\hat{J}\hat{K}} 
   S_{\mathcal{I}\hat{K}\hat{\alpha}}    \mathbb{H}^{-1}_{\hat{\alpha}\hat{\beta}}
   S_{\mathcal{J}\hat{\beta}\hat{J}}    
 \\ &
 - 8 \mathbb{A}^{-1}_{\hat{J}\hat{K}} S_{\mathcal{I}\hat{K}\bar{\eta}} 
   \mathbb{F}^{-1}_{\bar{\eta}\eta} S_{\mathcal{J}\eta \hat{J}} 
 + \mathbb{H}^{-1}_{\hat{\alpha}\hat{\sigma}} 
     S_{\mathcal{J}\hat{\sigma}\hat{\rho}}\,
     \mathbb{H}^{-1}_{\hat{\rho}\hat{\tau}}
	   S_{\mathcal{I}\hat{\tau}\hat{\beta}}
     - \mathbb{H}^{-1}_{\hat{\alpha}\hat{\beta}} 
     S_{\mathcal{J}\mathcal{I}\hat{\beta}\hat{\alpha}}
    \\ &
     + 2 \frac{S_{\mathcal{J}\bar{\eta}\eta}} {S_{\bar{\eta}\eta}} \frac{S_{\mathcal{I}\bar{\eta}\eta}} {S_{\bar{\eta}\eta}}
     - 2 \frac{S_{\mathcal{J}\mathcal{I}\bar{\eta}\eta}} {S_{\bar{\eta}\eta}} \,,
 \end{split}
\label{ZeroOGamma}
\end{equation}
where we have used identity (\ref{novertex}) to restrict nonghost indices $\hat{J},\hat{K}$. In Eqs.~(\ref{ZeroOGamma}) to (\ref{2AA}), all terms are evaluated at zero effective fields. The cancellation between irregular loops works as follows. First, we take the two terms:
\begin{eqnarray}\label{DoubleDev}
     \mathbb{A}^{-1}_{JK}
     S_{\mathcal{J}\mathcal{I}KJ}
     - 2\frac{S_{\mathcal{J}\mathcal{I}\bar{\eta}\eta}}{S_{\bar{\eta}\eta}}
     &=&
     \mathbb{A}^{-1}_{(JK)^{*}}
     S_{\mathcal{J}\mathcal{I}(KJ)^{*}}
     + 2\mathbb{A}^{-1}_{n\mathcal{A}}
     S_{\mathcal{J}\mathcal{I}\mathcal{A}n}
     - 2\frac{S_{\mathcal{J}\mathcal{I}\bar{\eta}\eta}}{S_{\bar{\eta}\eta}}\,.
 \label{cancellation1}
\end{eqnarray}
Similarly to the case of the first derivative, $\mathbb{A}^{-1}_{n\mathcal{A}}|_0$ is the irregular propagator (\ref{ProAn}), and identity (\ref{equalirregular}) implies $S_{\mathcal{J}\mathcal{I}\mathcal{A}n}=S_{\mathcal{J}\mathcal{I}\bar{\eta}\eta}$. Therefore, the last two terms in (\ref{DoubleDev}) cancel exactly between them. Second, in the first term of Eq.~(\ref{ZeroOGamma}) we extract the $\mathcal{A}n \times \mathcal{A}n $ loop, and compare with the penultimate term, obtaining
\begin{eqnarray}\label{MultiD}
     - 2\mathbb{A}^{-1}_{\mathcal{A}n}
     S_{\mathcal{J}n\mathcal{A}}
     \mathbb{A}^{-1}_{\mathcal{A}n}
     S_{\mathcal{I}n\mathcal{A}}
     + 2\frac{S_{\mathcal{J}\bar{\eta}\eta}}{S_{\bar{\eta}\eta}}\frac{S_{\mathcal{I}\bar{\eta}\eta}}{S_{\bar{\eta}\eta}}
     =0 \,.
\label{cancellation2}
\end{eqnarray}

The only possible irregular loops have been canceled in (\ref{cancellation1}) and (\ref{cancellation2}). In Eq.~(\ref{ZeroOGamma}) there remain irregular propagators, but always multiplied by one regular propagator. In the first term, consider the possibility of having one $\mathbb{A}^{-1}_{\mathcal{A}\mathcal{A}}|_0$ propagator. In this case the vertex restriction $S_{\mathcal{I} \mathcal{A} \mathcal{A}} = 0$ forces the loop to be of the form
\begin{equation}
 \mathbb{A}^{-1}_{\mathcal{A}\mathcal{A}} S_{\mathcal{J}\mathcal{A}\hat{N}} 
 \mathbb{A}^{-1}_{\hat{N}\hat{K}} S_{\mathcal{I}\hat{K}\mathcal{A}} \,,
\label{2AA}
\end{equation}
the factor $\mathbb{A}^{-1}_{\hat{N}\hat{K}}$ being the regular propagator. Other possibility in the first term of (\ref{ZeroOGamma}) is one irregular propagator in the form $\mathbb{A}^{-1}_{\mathcal{A}n}
S_{\mathcal{J}nN} \mathbb{A}^{-1}_{(NK)^*} S_{\mathcal{I}K\mathcal{A}}$, since in (\ref{cancellation2}) we have extracted the $\mathcal{A}n \times \mathcal{A}n$ combination. The factor $\mathbb{A}^{-1}_{(NK)^*}$ is a regular propagator. Due again to $S_{\mathcal{I} \mathcal{A} \mathcal{A}} = 0$, the irregular propagator $\mathbb{A}^{-1}_{\mathcal{A}\mathcal{A}}$ cannot arise in the second term of (\ref{ZeroOGamma}). The last irregular propagator in (\ref{ZeroOGamma}) is the factor $\mathbb{F}^{-1}_{\bar{\eta}\eta}$ in the fourth term. In this case the factor $\mathbb{A}^{-1}_{\hat{J}\hat{K}}$ is the regular propagator closing the loop.

\subsubsection{Generalization to derivatives of arbitrary order}
We recall our index conventions: indices $\mathcal{I},\mathcal{J},\ldots$ run for all fields, whereas $I,J,\ldots$ run for nonghost bosonic fields. Index $\hat{I}$ is the same as $I$, excluding $I = \mathcal{A}$. $\alpha,\beta,\ldots$ run for all ghosts fields, whereas $\hat{\alpha},\hat{\beta},\ldots$ run for BFV ghosts only. The only irregular propagators of the theory are $\left<\mathcal{A}\mathcal{A}\right>$, $\left<\mathcal{A}n\right>$, and $\left< \bar{\eta}\eta \right>$. With this notation, all entries $\mathbb{A}^{-1}_{\hat{I}\hat{J}}|_0$ and $\mathbb{H}^{-1}_{\hat{\alpha}\hat{\beta}}|_0$ are strictly regular propagators. 
	
We collect all the identities restricting the vertices, which are due to the dependence of the classical action $S$ on the fields $\mathcal{A}$, $\eta,\bar{\eta}$ and the BFV ghosts. $S$ is linear in $\mathcal{A}$, bilinear in ghost fields $\eta,\bar{\eta}$, and there are no mixed terms between $\mathcal{A}$, the BFV ghosts, and the $\eta,\bar{\eta}$ ghost. Hence, for an arbitrary number of derivatives, we can establish the identities:
\begin{equation}
	S_{\mathcal{I}\cdots\mathcal{J} \mathcal{A} \mathcal{A} } 
	= 
	S_{\mathcal{I}\cdots\mathcal{J} \mathcal{A} \alpha } 
	=
	S_{\mathcal{I}\cdots\mathcal{J} \eta \eta }
	=
	S_{\mathcal{I}\cdots\mathcal{J} \bar{\eta} \bar{\eta} }
	=
	S_{\mathcal{I}\cdots\mathcal{J} \eta \hat{\alpha}}
	=
	S_{\mathcal{I}\cdots\mathcal{J} \bar{\eta} \hat{\alpha}} 
	= 
	0 \,.
	\label{novertexA}
\end{equation}
Due to relation (\ref{equalirregular}), we have the general identity between derivatives
\begin{equation}
S_{\mathcal{I} \cdots \mathcal{J} n \mathcal{A} }
=
S_{\mathcal{I} \cdots \mathcal{J} \bar{\eta} \eta } \,,
\label{equalderivative}
\end{equation}
which implies the equality between irregular propagators $\mathbb{A}^{-1}_{\mathcal{A} n}|_0 = (S_{\bar{\eta}\eta})^{-1}|_0$.

According to Eq.~(\ref{Berfull}), the one-loop quantum correction of the effective action is the functional \footnote{In the remainder of the paper we omit the factor $-2i$, the bar and the $^{(1)}$ symbol in $\Gamma$.}
	\begin{equation}
	\Gamma = 
	\ln \left( \frac{ \det \mathbb{M} }{ \det \mathbb{D} } \right)
	=
	\ln \det \mathbb{M} - \ln \det \mathbb{D} 
	\equiv \Gamma^{(\mathbb{M})} - \Gamma^{(\mathbb{D})}
	\,,
	\label{Gamma}
	\end{equation}
where 
	\begin{equation}
	\mathbb{M} = \mathbb{A} - 2\mathbb{B} \mathbb{D}^{-1} \mathbb{B}^T \,.
	\end{equation}
Due to expression (\ref{Gamma}), derivatives of $\Gamma$ have the same functional dependence on $\mathbb{M}$ and $\mathbb{D}$, with the corresponding change of sign.
	
We can obtain a formula representing the arbitrary derivative of the functionals $\Gamma^{(\mathbb{M})}$ and $\Gamma^{(\mathbb{D})}$. As a consequence of the formulas
\begin{equation}
	\frac{ \delta \ln \det \mathbb{M} }{ \delta \Phi^{\mathcal{I}} } =
	\mathbb{M}^{-1}_{IJ}
	\frac{ \delta \mathbb{M}_{JI} }{ \delta \Phi^{\mathcal{I}} } \,,
	\qquad
	\frac{ \delta \mathbb{M}^{-1}_{IJ} }{ \delta \Phi^{\mathcal{I}} } =
	- \mathbb{M}^{-1}_{IK}  
	\frac{ \delta \mathbb{M}_{KL} }{ \delta \Phi^{\mathcal{I}} } 
	\mathbb{M}^{-1}_{LJ} \,,
\end{equation}
the $N$-order derivative of $\Gamma^{(\mathbb{M})}$ can be written as a sum of terms that are multiplications of factors with the same generic structure: each factor is an inverse $\mathbb{M}^{-1}$ times a derivative of $\mathbb{M}$ of certain order. Consider a given order $N$ in derivatives. We introduce the set of partitions of the number $N$, parameterized by the independent integers $\{ n_1 , \ldots , n_N \}$, each one taking values in $n_i= 0,1,\ldots,N$, and subject to the global condition
	\begin{equation}
	n_1 + n_2 + \cdots n_N = N \,.
	\label{N}
	\end{equation}
We denote an arbitrary partition $\{ n_i \}$ satisfying (\ref{N}) as $P[n_k]$. The $N$-order derivative of $\Gamma^{(\mathbb{M})}$ can be represented by the formula
	\begin{equation}
	\begin{split}
	&
	\frac{ \delta^N \Gamma^{(\mathbb{M})} }
	{ \delta \Phi^{\mathcal{I}_1} \cdots \delta \Phi^{\mathcal{I}_N } } 
	= 
	\\ &
	\sum\limits_{P[n_k]} 
	\sum\limits_{ \substack{
			\mathcal{J},\ldots,\mathcal{L} \,\subseteq \mathcal{I}
			\\[.5ex]
			\mathcal{J} \cup \mathcal{K} \cdots \cup \mathcal{L} 
			= \mathcal{I}
			\\[.5ex]
			\mathcal{J} \cap \mathcal{K} = \emptyset\,,\ldots
	} }
	C_P \,
	\mathbb{M}^{-1}_{IJ} 
	\frac{ \delta^{n_1} \mathbb{M}_{JK} }
	{ \delta \Phi^{\mathcal{J}_1} \cdots \delta \Phi^{\mathcal{J}_{n_1}} }
	\mathbb{M}^{-1}_{KL} 
	\frac{ \delta^{n_2} \mathbb{M}_{LM} }
	{ \delta \Phi^{\mathcal{K}_1} \cdots \delta \Phi^{\mathcal{K}_{n_2}} }
	\cdots 
	\mathbb{M}^{-1}_{PQ} 
	\frac{ \delta^{n_N} \mathbb{M}_{QI} }
	{ \delta \Phi^{\mathcal{L}_1} \cdots \delta \Phi^{\mathcal{L}_{n_N}} } 
	\end{split}
	\,.
	\label{dNGamma}
	\end{equation}
In the first summation $\sum_{P[n_k]}$, each term corresponds to an specific partition $P[n_k]$. In the second summation, we use multi-index notation: $\mathcal{I} = \{ \mathcal{I}_1,\ldots,\mathcal{I}_N\}$, and $\mathcal{J} = \{ \mathcal{J}_1,\ldots \mathcal{J}_{n_1} \}$, $\mathcal{K} = \{ \mathcal{K}_1,\ldots \mathcal{K}_{n_2} \}$,...,$\mathcal{L} = \{ \mathcal{L}_1,\ldots \mathcal{L}_{n_N} \}$. For each term of the second summation, the given derivatives $\mathcal{I}$ must be distributed among the subsets $\mathcal{J},\mathcal{K},\ldots$ without repeating indices between the subsets. If $n_i = 0$, the whole factor $\mathbb{M}^{-1}_{IJ} \delta^{0} \mathbb{M}_{JK}$ is replaced by $\delta_{IK}$. $C_P$ is the coefficient (with sign) of the given term. When evaluated at zero effective fields, $\mathbb{M}^{-1}_{IJ}$ gives the propagators and $ \delta^{n_i} \mathbb{M}_{JK}$ gives the vertices, joined by the legs in the index $J$. Each term in the entire summation forms a loop. The index $I$ of the last factor is contracted with the index $I$ of the first one to close the loop. We leave unspecified the coefficient $C_P$. We do not require to know it explicitly, since the functional dependence of $\Gamma$ is the same on $\mathbb{M}$ and $\mathbb{D}$, with opposite sign. Hence, for any term with derivatives of $\mathbb{M}$, there is a counter-term with derivatives of $\mathbb{D}$ with the same coefficient and opposite sign. This is what we require to obtain the cancellations.
	
Since $\mathbb{B}$ has no zeroth order term, the evaluation of the derivative (\ref{dNGamma}) at zero effective fields results
\begin{equation}
	\begin{split}
	&
	\left. \frac{ \delta^N \Gamma^{(\mathbb{M})} }
	{ \delta \Phi^{\mathcal{I}_1} \cdots \delta \Phi^{\mathcal{I}_N} } 
	\right|_0 =  
	\\ & 
	\left. 
	\sum\limits_{P[n_k]} 
	\sum\limits_{ \mathcal{J}\ldots }
	C_P 
	\mathbb{A}^{-1}_{IJ} 
	\frac{ \delta^{n_1}
		\left[ \mathbb{A}_{JK} - 2( \mathbb{B} \mathbb{D}^{-1} \mathbb{B}^T )_{JK} \right] }
	{ \delta \Phi^{\mathcal{J}_1} \cdots \delta \Phi^{\mathcal{J}_{n_1}} }
	\cdots 
	\mathbb{A}^{-1}_{PQ} 
	\frac{ \delta^{n_N}
		\left[ \mathbb{A}_{QI} - 2( \mathbb{B} \mathbb{D}^{-1} \mathbb{B}^T )_{QI} \right] }
	{ \delta \Phi^{\mathcal{L}_1} \cdots \delta \Phi^{\mathcal{L}_{n_N}} }
	\right|_0 \,.
	\end{split}
	\label{expansion}
\end{equation}
Let us analyze first the terms that contain only derivatives of $\mathbb{A}$, 
\begin{eqnarray}
	&&
	\left. 
	\sum\limits_{P[n_k]} 
	\sum\limits_{ \mathcal{J}\ldots }
	C_P 
	\mathbb{A}^{-1}_{IJ} 
	\frac{ \delta^{n_1} \mathbb{A}_{JK} }
	{ \delta \Phi^{\mathcal{J}_1} \cdots \delta \Phi^{\mathcal{J}_{n_1}} } 
	\cdots
	\mathbb{A}^{-1}_{PQ} 
	\frac{ \delta^{n_N} \mathbb{A}_{QI} }
	{ \delta \Phi^{\mathcal{L}_1} \cdots \delta \Phi^{\mathcal{L}_{n_N}} } 
	\right|_0 
	\nonumber
	\\
	&& =
	\left. 
	\sum\limits_{P[n_k]} 
	\sum\limits_{ \mathcal{J}\ldots }
	C_P 
	\mathbb{A}^{-1}_{IJ} S_{\mathcal{J}_1 \cdots \mathcal{J}_{n_1} JK}
	\cdots
	\mathbb{A}^{-1}_{PQ} S_{\mathcal{L}_1 \cdots \mathcal{L}_{n_N} QI}
	\right|_0 \,, 
	\label{AdA}
	\end{eqnarray}
leaving for later the factor $\mathbb{B} \mathbb{D}^{-1} \mathbb{B}^T$. Our first task is to prove that if one irregular propagator $\left< \mathcal{A}\mathcal{A} \right> = \mathbb{A}^{-1}_{\mathcal{A}\mathcal{A}}|_0$ arises in (\ref{AdA}), then it is accompanied by at least one regular propagator $\mathbb{A}^{-1}_{\hat{I}\hat{J}}|_0$. Suppose that one propagator $\mathbb{A}^{-1}_{\mathcal{A}\mathcal{A}}|_0$ arises in the last factor of (\ref{AdA}), $PQ = \mathcal{A}\mathcal{A}$. Then, this term has the form
\begin{equation}
	\left.
	\mathbb{A}^{-1}_{\hat{I}J} 
	S_{\mathcal{J}_1 \cdots \mathcal{J}_{n_1} J K }
	\cdots
	\mathbb{A}^{-1}_{LH} 
	S_{\mathcal{M}_1 \cdots \mathcal{M}_{n_{N-2}} H M }
	\mathbb{A}^{-1}_{M\hat{N}} 
	S_{\mathcal{K}_1 \cdots \mathcal{K}_{n_{N-1}} \hat{N} \mathcal{A}}
	\mathbb{A}^{-1}_{\mathcal{A}\mathcal{A}} 
	S_{\mathcal{L}_1 \cdots \mathcal{L}_{n_N} \mathcal{A} \hat{I} }  \right|_0 \,,
	\label{AA}
	\end{equation}
where we have used conditions (\ref{novertexA}) to restrict $N\rightarrow\hat{N}$ and $I\rightarrow\hat{I}$. Index $M$ is not restricted, hence it may take the value $M = \mathcal{A}$, producing an irregular propagator $\mathbb{A}^{-1}_{\mathcal{A} n}$. But in this case the next vertex also gets the leg $M=\mathcal{A}$, hence restricting $H\rightarrow\hat{H}$ by (\ref{novertexA}). We can continue iteratively the contractions of indices, until we reach the first factor with $K = \mathcal{A}$ and $J \rightarrow \hat{J}$. We then obtain that the first factor $\mathbb{A}^{-1}$ is the regular propagator $\mathbb{A}^{-1}_{\hat{I}\hat{J}}|_0$. Therefore, irregular loops contained in (\ref{AdA}) cannot have $\left<\mathcal{A}\mathcal{A} \right>$ propagators. 
	
Since the other irregular propagator in the bosonic sector is $\left< \mathcal{A} n\right> = \mathbb{A}^{-1}_{\mathcal{A} n}|_0$, we have that the only possibility to form an irregular loop from (\ref{AdA}) is a loop made completely of $\left< \mathcal{A} n \right>$ propagators. The sum of all irregular loops of this kind is given by
\begin{equation}
	\left.
	\sum\limits_{P[n_k]}
	\sum\limits_{ \mathcal{J}\ldots }
	C_P 
	2 \mathbb{A}^{-1}_{\mathcal{A} n} 
	S_{\mathcal{J}_1 \cdots \mathcal{J}_{n_1} n \mathcal{A} }
	\cdots
	2 \mathbb{A}^{-1}_{\mathcal{A} n} 
	S_{\mathcal{L}_1 \cdots \mathcal{L}_{n_N} n \mathcal{A} }
	\right|_0 \,.
	\label{irregularAn}
\end{equation}
Factors of $2$ arise due to the two possibilities $IJ = \mathcal{A} n , n \mathcal{A}$ in each factor. On the other hand, the $N$-order derivative of $\Gamma^{(\mathbb{D})}$ is
\begin{eqnarray}
	\left.
	\frac{ \delta^N \Gamma^{(\mathbb{D})} }
	{ \delta \Phi^{\mathcal{I}_1} \cdots \delta \Phi^{\mathcal{I}_N} } \right|_0 
	&=& 
	\left.
	\sum\limits_{P[n_k]} 
	\sum\limits_{ \mathcal{J}\ldots }
	C_P 
	\mathbb{D}^{-1}_{\alpha\beta} 
	\frac{ \delta^{n_1} \mathbb{D}_{\beta\gamma} }
	{ \delta \Phi^{\mathcal{J}_1} \cdots \delta \Phi^{\mathcal{J}_{n_1}} } 
	\cdots
	\mathbb{D}^{-1}_{\rho\sigma} 
	\frac{ \delta^{n_N} \mathbb{D}_{\sigma\alpha} }
	{ \delta \Phi^{\mathcal{L}_1} \cdots \delta \Phi^{\mathcal{L}_{n_N}} }
	\right|_0 
	\nonumber
	\\
	&=&
	\left.
	\sum\limits_{P[n_k]} 
	\sum\limits_{ \mathcal{J}\ldots }
	C_P 
	\mathbb{D}^{-1}_{\alpha\beta} 
	S_{\mathcal{J}_1 \cdots \mathcal{J}_{n_1} \beta \gamma} 
	\cdots
	\mathbb{D}^{-1}_{\rho\sigma} 
	S_{\mathcal{L}_1 \cdots \mathcal{L}_{n_N} \sigma \alpha}
	\right|_0.
 \label{irregularghostsBFVSC}
\end{eqnarray}
In the ghost sector, the only irregular propagator is $\left< \bar{\eta}\eta \right> = \mathbb{D}^{-1}_{\bar{\eta}\eta}$. Due to identities (\ref{novertexA}), if one element $\mathbb{D}^{-1}$ takes the entry $\bar{\eta}\eta$, then the rest of propagators in (\ref{irregularghostsBFVSC}) necessarily take the same entry (or the transpose), forming an irregular loop. There is no mixing between the propagator of $\bar{\eta}\eta$ and the propagators of the BFV ghosts. From (\ref{irregularghostsBFVSC}), let us write apart the sum of all  irregular loops with the $\bar{\eta}\eta$ propagators. It is equal to
\begin{eqnarray}
	\left.
	\sum\limits_{P[n_k]} 
	\sum\limits_{ \mathcal{J}\ldots }
	C_P 
	2 (S_{\bar{\eta}\eta})^{-1} 
	S_{\mathcal{J}_1 \cdots \mathcal{J}_{n_1} \eta \bar{\eta}} 
	\cdots
	2 (S_{\bar{\eta}\eta})^{-1} 
	S_{\mathcal{L}_1 \cdots \mathcal{L}_{n_N} \eta \bar{\eta}}
	\right|_0    \,.
	\label{irregularghosts}
	\end{eqnarray}
Again, factors of $2$ come from the symmetry between the two possibilities $\bar{\eta}\eta , \eta\bar{\eta}$. The other terms in (\ref{irregularghostsBFVSC}) correspond to regular loops made with the BFV ghost propagators. Identity (\ref{equalderivative}), which also equals the irregular propagators, implies that all irregular loops in (\ref{irregularAn}) and (\ref{irregularghosts}) cancel between them exactly. Note that the cancellations are one-to-one: for each irregular loop with $\left< \mathcal{A} n \right>$ in (\ref{irregularAn}), there is an irregular loop with $\left< \bar{\eta} \eta \right>$ in (\ref{irregularghosts}), with the same coefficient.
	
Finally, we analyze the factor
\begin{equation}
	\left.
	\mathbb{A}^{-1}_{IJ} 
	\frac{ \delta^{n_i} 
		\left( \mathbb{B} \mathbb{D}^{-1} \mathbb{B}^T \right)_{JK} }
	{ \delta \Phi^{\mathcal{J}_1} \cdots \delta \Phi^{\mathcal{J}_{n_i}} } 
	\right|_0 
	=
	\left.
	\mathbb{A}^{-1}_{I\hat{J}} 
	\frac{ \delta^{n_i}
		\left( \mathbb{B}_{\hat{J}\alpha} \mathbb{D}^{-1}_{\alpha\beta} \mathbb{B}^T_{\beta \hat{K}} \right) }
	{ \delta \Phi^{\mathcal{J}_1} \cdots \delta \Phi^{\mathcal{J}_{n_i}} } 
	\right|_0 \,,
	\label{BDC}
\end{equation}
that we omitted previously in (\ref{expansion}). In (\ref{BDC}) we have used conditions (\ref{novertexA}) to restrict indices $\hat{J}$ and $\hat{K}$. Since $\mathbb{B}$ starts at linear order, the only terms that survive when evaluated at zero are those for which at least one derivative falls in $\mathbb{B}$ and at least one derivative falls in $\mathbb{B}^T$. At zero fields, these derivatives are vertices of the form $S_{\mathcal{P} \cdots \mathcal{Q} \hat{J} \alpha}$ and $S_{\mathcal{T} \cdots \mathcal{U} \beta \hat{K}}$, respectively, where $\mathcal{P}\cdots\mathcal{Q}$, $\mathcal{T}\cdots\mathcal{U}$ are subsets of the derivatives $\mathcal{J}_{1}\cdots\mathcal{J}_{n_i}$. We can write a generic nonzero term coming from (\ref{BDC}) as
\begin{equation}
	\left.
	\mathbb{A}^{-1}_{I\hat{J}} 
	S_{\mathcal{P} \cdots \mathcal{Q} \hat{J} \alpha} 
	\frac{   \delta^p 
		\left(\mathbb{D}^{-1}_{\alpha\beta} \right) }
	{ \delta \Phi^{\mathcal{R}} \cdots \delta \Phi^{\mathcal{S}} } 
	S_{\mathcal{T} \cdots \mathcal{U} \beta \hat{K}}
	\right|_0  \,.
	\label{crossed}
\end{equation}
Now we want to prove that, when this kind of factor arises in one term of (\ref{expansion}), it is  necessarily accompanied by at least one regular propagator $\mathbb{A}^{-1}_{\hat{I}\hat{J}}|_0$. The reasoning is similar to the iterative procedure we have done previously. If the factor $\mathbb{A}^{-1}_{I\hat{J}}$ written in (\ref{crossed}) is an irregular propagator due to $I = \mathcal{A}$, we continue the index contractions iteratively passing over irregular propagators. The iterations continue until we encounter an additional factor of the form (\ref{crossed}); hence with a vertex $S_{\mathcal{K} \cdots \mathcal{L} \beta \hat{K}}$, or directly the same factor (\ref{crossed}) if we turn around the loop. In both cases the index $I$ of the last propagator $\mathbb{A}^{-1}$ is restricted by the contraction with the corresponding index $\hat{K}$ of $S_{\mathcal{K} \cdots \mathcal{L} \beta \hat{K}}$; hence $I \rightarrow \hat{I}$. Therefore, this last propagator is the regular propagator $\mathbb{A}^{-1}_{\hat{I}\hat{J}}|_0$. As a consequence, no irregular loops can be formed in a term of (\ref{expansion}) if at least one factor (\ref{BDC}) is present in the term. The only irregular loops in the derivative (\ref{expansion}) has been considered in (\ref{irregularAn}) and (\ref{irregularghosts}), and they cancel each other exactly.


\section*{Conclusions}
We have found an integrated form of the one-loop effective action of the Ho\v{r}ava theory, in its more general nonprojectable version, incorporating the second-class constraints to the effective action. The complete quantized theory we have taken is the $(2+1)$-dimensional theory. We remark that ghost fields allow us to promote the measure of the second-class constraints to the quantum canonical Lagrangian, and these ghost can be used to determine the form of the effective action in a similar way as the ghosts associated with the gauge symmetry. The quantization is based on the BFV formalism, which allows us to use the required gauge-fixing condition, analogous to the projectable case, to obtain regular propagators on most of the fields. The result for the effective action has a quite compact form; it is the Berezinian of the matrix of second order in derivatives, where all ghost fields enter in the odd sector of the Berezinian. Thus, in this sense one can handle the second-class constraints with no more obstacles than the first-class constraint.

We have shown that the irregular loops cancel exactly between them in the effective action. We have given a general proof for any order in derivatives of the action. Our analysis shows how the loops formed with the ghosts of the second-class constraints located in the denominator of the Berezinian cancel the sector of nonghost irregular loops. Hence, it becomes clear the role of these ghosts in eliminating dangerous degrees of freedom in the loop expansion. This is similar to the role of the BFV and Faddeev-Popov ghosts in eliminating gauge degrees of freedom. In the remaining loops, there always arises at least one regular propagator. The regular propagator is sufficient to render the loop regular, in the sense that the integration of the frequency $\omega$ is unaltered by the irregular loops. Thus, the integration on $\omega$ and $k^i$ must be analyzed on the same grounds as if the theory was not affected by irregular propagators.

The cancellation of the irregular loops in the effective action is an essential step toward the renormalization of the theory. Once the divergences coming from the irregular loops have been eliminated, the analysis can be concentrated on the divergences coming from the regular loops, as in the case of the projectable theory \cite{Barvinsky:2015kil,Barvinsky:2017zlx}. We expect that the form of the effective action we have obtained can be helpful for this purpose.

There exist other versions of the Ho\v{r}ava theory for which we may expect that the BFV quantization and the computation of the effective action can be applied as well. In the side of the nonprojectable version there is, for example, a special case, called the critical case, which has the extra scalar mode eliminated \cite{Bellorin:2013zbp,Bellorin:2016wsl}. There is also a well-known version with an additional $U(1)$ symmetry \cite{Zhu:2011xe}, that also eliminates the extra mode (the $U(1)$ symmetry was introduced in the projectable version in Ref.~\cite{Horava:2010zj}). To apply the BFV quantization to these theories, it is required to take the algebra of first- and second-class constraints  (see Ref.~\cite{Mukohyama:2015gia}) and to analyze its consequences on the definition of the measure of the path integral. In the case of the version with the $U(1)$ symmetry, more BFV ghosts should arise due to the extra gauge symmetry.




\end{document}